\documentclass{mn2e}
\usepackage{epsf,times}
\newcommand{\etal}{{et al}\/.}

\begin{document}
\title[X-ray emission from Pictor A]{The {\it Chandra} view of extended X-ray emission from Pictor A}
\author[M.J.~Hardcastle \& J.H.~Croston]{M.J.\ Hardcastle$^1$ and
  J.H.\ Croston$^2$\\$^1$ School of Physics,
  Astronomy and Mathematics, University of
Hertfordshire, College Lane, Hatfield, Hertfordshire AL10 9AB\\
$^2$ Service d'Astrophysique, CEA Saclay, L'Orme des Merisiers,
  91191 Gif-sur-Yvette, France}
\maketitle
\begin{abstract}
We discuss the extended X-ray emission seen in three archival {\it
  Chandra} observations, and one archival {\it XMM-Newton}
  observation, of the FRII radio galaxy Pictor A. The overall
  properties of the X-ray lobes are consistent with the conclusions of
  earlier workers that the extended X-ray emission is largely due to
  the inverse-Compton process, and the implied departure from
  equipartition is in the range seen by us in other sources. In
  detail, we show that the X-ray/radio flux ratio varies quite
  strongly as a function of position throughout the source, and we
  discuss possible implications of this observation for the spatial
  variation of electron energy spectra and magnetic field strength
  through the lobe. We show that the radio and X-ray properties of the
  lobe are not consistent with a simple model in which variations in
  the magnetic field strength alone are responsible for the observed
  differences between emission at different frequencies. We also
  discuss the origins of the extended emission seen around the eastern
  hotspot, arguing that it may be diffuse synchrotron radiation
  tracing a region of distributed particle acceleration, and the
  implications of a possible weak X-ray counterjet detection which,
  taken together with the other properties of the bright X-ray jet,
  leads us to suggest that the X-ray jet and possible counterjet are
  also produced by synchrotron emission.
\end{abstract}

\section{Introduction}

Lobe-related X-ray emission has now been seen in a large number of
powerful (FRII: Fanaroff \& Riley 1974) radio galaxies and quasars
(Kataoka \& Stawarz 2005; Croston \etal\ 2005). Although in a few
cases the X-ray emission coincident with the lobes has been
interpreted in terms of shocked thermal material (e.g. Nulsen \etal\
2005) it seems most likely, from the good agreement between the
observed flux levels and the predictions of an inverse-Compton model,
that the vast majority of sources are dominated by inverse-Compton
emission. This unexpected result has provided us with valuable
insights into the energetics and particle content of FRII lobes; for
example, it severely reduces the likelihood of an energetically
dominant proton population (Croston \etal\ 2005) and where comparisons
have been possible (e.g. Hardcastle \etal\ 2002, Croston \etal\ 2004)
it suggests that the internal pressure of the lobes is comparable to
the pressure in the external thermal medium.

The cosmic microwave background generally provides the energetically
dominant photon population in the lobes of large radio sources,
although infrared photons from the active nucleus may be important in
some cases (e.g. Brunetti, Setti \& Comastri 1997). In either case,
since relatively high-energy photons are being scattered into the
X-ray, the inverse-Compton emission tells us about the low-energy
electron population, with Lorentz factors $\gamma \sim 10^3$ for
scattering of the cosmic microwave background (CMB) and $\sim 30$ for
nuclear infrared photons. The synchrotron emission from these
electrons is generally unobservable with the current generation of
radio telescopes, appearing at tens of MHz or below (e.g. Harris
2004). Thus, if we have identifiable inverse-Compton emission from a
population of electrons whose energies can be inferred by other means,
the X-ray data can give us spatially resolved information about the
low-energy electron population that can currently be obtained in no
other way.

To do this in practice requires good statistics in the X-ray, and this
rules out many distant FRII sources, where the lobe X-ray detection
may consist of only a few tens of counts, sufficient to estimate the
overall lobe magnetic field strength (e.g. Croston \etal\ 2005) but
not to look in detail at spatial variations of the X-ray surface
brightness as a function of radio brightness. Some work has been done
in this area with long observations (e.g. Isobe \etal\ 2002, 2005) but it
has not been possible to draw very strong conclusions about the
low-energy electron population because the number of photons collected
is still small for most sources. For a fixed lobe magnetic field
strength, inverse-Compton lobe emissivity scales linearly with
synchrotron emissivity, and so the brightest radio sources are a good
place to start searching for bright inverse-Compton emission. However,
the X-ray emission from several of the brightest extragalactic radio
sources in the sky (Cygnus A, Cen A, M87, Her A) is dominated by
thermal bremsstrahlung from the cluster environment, while the lobes
of Fornax A, which provided the original claimed detection of lobe
inverse-Compton emission (Feigelson \etal\ 1995, Kaneda \etal\ 1995,
Tashiro \etal\ 2001) and `Centaurus B' (PKS B1343$-$601: Tashiro
\etal\ 1998) are too large for observations with the current
generation of X-ray telescopes to be very useful. However, Pictor A,
roughly the 8th brightest radio source in the sky in terms of flux
density at low frequencies [the list of Robertson (1973) omits Cyg A
and Cen B] has little thermal emission, is well matched in size (LAS 8
arcmin) to the {\it Chandra} field of view, and is an FRII source,
which allows a comparison with the large number of more distant
powerful FRIIs observed to have lobe X-ray emission (Kataoka \&
Stawarz 2005; Croston \etal\ 2005). The original {\it Chandra}
observation of Pic A was discussed by Wilson, Young \& Shopbell (2001:
hereafter W01). In addition to X-ray detections of a bright jet and of
the previously known X-ray hotspot, they noted some extended X-ray
emission associated with the E lobe. Grandi \etal\ (2003) used {\it
XMM-Newton} observations of the source to show that extended emission
from both lobes was detected and to argue for an inverse-Compton
origin for the extended X-ray emission, but were only able to discuss
in detail the X-rays from a small region near the E hotspot. In this
paper we revisit the {\it Chandra} data of W01 and the {\it
XMM-Newton} data of Grandi \etal\ (2003), together with new data
available in the {\it Chandra} archive, and use the spatial resolution
of {\it Chandra} and the sensitivity provided by more than 110 ks of
observations to discuss the relationship between the large-scale radio
and X-ray emission in the source.

Pic A is a broad-line radio galaxy with $z=0.035$ (e.g. Simkin \etal\
1999). In what follows we use a concordance cosmology with $H_0 = 70$
km s$^{-1}$ Mpc$^{-1}$, $\Omega_{\rm m} = 0.3$ and $\Omega_\Lambda =
0.7$; at the redshift of Pic A, 1 arcsec corresponds to 700 pc. All
spectral fits include Galactic absorption with a column density of
$4.2 \times 10^{20}$ cm. Spectral indices $\alpha$ are the energy
indices and are defined in the sense that flux $\propto \nu^{-\alpha}$.
Where a spectral index is quoted with subscript and superscript
frequency values, it represents the two-point radio spectral index
between the specified frequencies.

\section{Observations and analysis}

\subsection{Chandra}

In addition to some short exploratory observations, Pic A has been
observed three times with {\it Chandra} (see Table \ref{obs} for
details). The first observation, reported by W01, was centred on the
active nucleus, and the lobes span the ACIS-S2 and S3 chips; the
pointing for the two later observations was close to the western
hotspot, and much of the E lobe was undetectable, falling between the
ACIS-S and I arrays, although a small part of the extreme eastern end
falls on the ACIS-I3 chip (Fig. \ref{rawims}). 

All three long observations were reprocessed from the level 1 events
files using the latest version of {\sc ciao} at the time of writing
(3.2.1) and the corresponding {\sc caldb} (3.0.1). The latest gain
files were applied, the S4 chip destreaked, and the frame transfer
streaks removed (without background replacement) using standard
techniques detailed in the {\sc ciao} on-line
documentation\footnote{http://asc.harvard.edu/ciao/}. For the first
long observation there were some significant periods of high
background, and so we filtered the data using the results of the {\it
analyze\_ltcrv} script, giving the effective live time listed in Table
\ref{obs}. For the other two observations we only applied the standard
time filtering.

The spectral extractions described in the rest of the paper used the
{\sc ciao} script {\it acisspec} to extract spectra and suitably
weighted response files for extended regions. For the 2002
observations we generated new response files using the {\it mkacisrmf}
tools.

\subsection{XMM-Newton}

In order to confirm our {\it Chandra} results, and to extend the
results of Grandi et al. (2003) to larger lobe regions comparable with
our {\it Chandra} analysis, we extracted the archive {\it XMM-Newton}
observation of Pictor A (obsID = 0090050701) described by Grandi et
al. (2003). The observation, obtained on 2001 March 17, had a livetime
of 20521 ks. The MOS cameras were in small-window mode, so that the
lobe regions were not observed; we therefore used only the pn
data. The data were reprocessed using the {\it XMM-Newton} SAS version
6.0.0, and the latest calibration files from the {\it XMM-Newton}
website. The pn data were filtered to include only single and double
events (PATTERN $\leq 4$), and the standard flag filter \#XMMEA\_EP (but
excluding bad columns and rows). A histogram in the energy range 12.0
- 14.0 keV was used to look for background flaring. There was
significant flaring activity throughout the observation, and so we
excluded all periods with a count rate higher than 2 cts s$^{-1}$,
giving a filtered dataset of duration 10~062 s. This count level was
chosen by eye so as to cut out the worst flares above the quiescent
background ($\sim 0.5$ cts s$^{-1}$) while not excluding too large a
fraction of the dataset. Grandi \etal\ did not
gti-filter the data for their analysis, but we found that the
filtering we applied did not significantly limit our scientific analysis.

\subsection{Radio}

For comparison between the radio and X-ray structures and analysis of
the radio spectrum we use the radio data of Perley, R\"oser \&
Meisenheimer (1997: hereafter P97), kindly provided by Rick Perley.
Maps at central frequencies of 327.5 MHz, 1.471 GHz and 4.847 GHz are
used in the analysis -- for simplicity in what follows we refer to
these as the 330-MHz, 1.5-GHz and 4.8-GHz maps. The images we use are
those in which P97 used a circular restoring beam.

\begin{table*}
\caption{The {\it Chandra} observations of Pictor A used in this
  paper}
\label{obs}
\begin{tabular}{lrllrr}
\hline
Date&OBSID&\multicolumn{2}{c}{Pointing position}&Livetime&Livetime after
\\
&&RA (h m s)&Dec ($^\circ$ $'$ $''$)&(s)&filtering (s)\\
\hline
2000 Jan 18&346&05:19:49.7&$-$45:46:45.00&25737&25737\\
2002 Sep 17&3090&05:19:26.2&$-$45:45:53.50&46959&36351\\
2002 Sep 22&4369&05:19:26.2&$-$45:45:53.50&49126&49126\\
\hline
\end{tabular}
\end{table*}

\begin{figure*}
\hbox{
\epsfxsize 7.5cm
\epsfbox{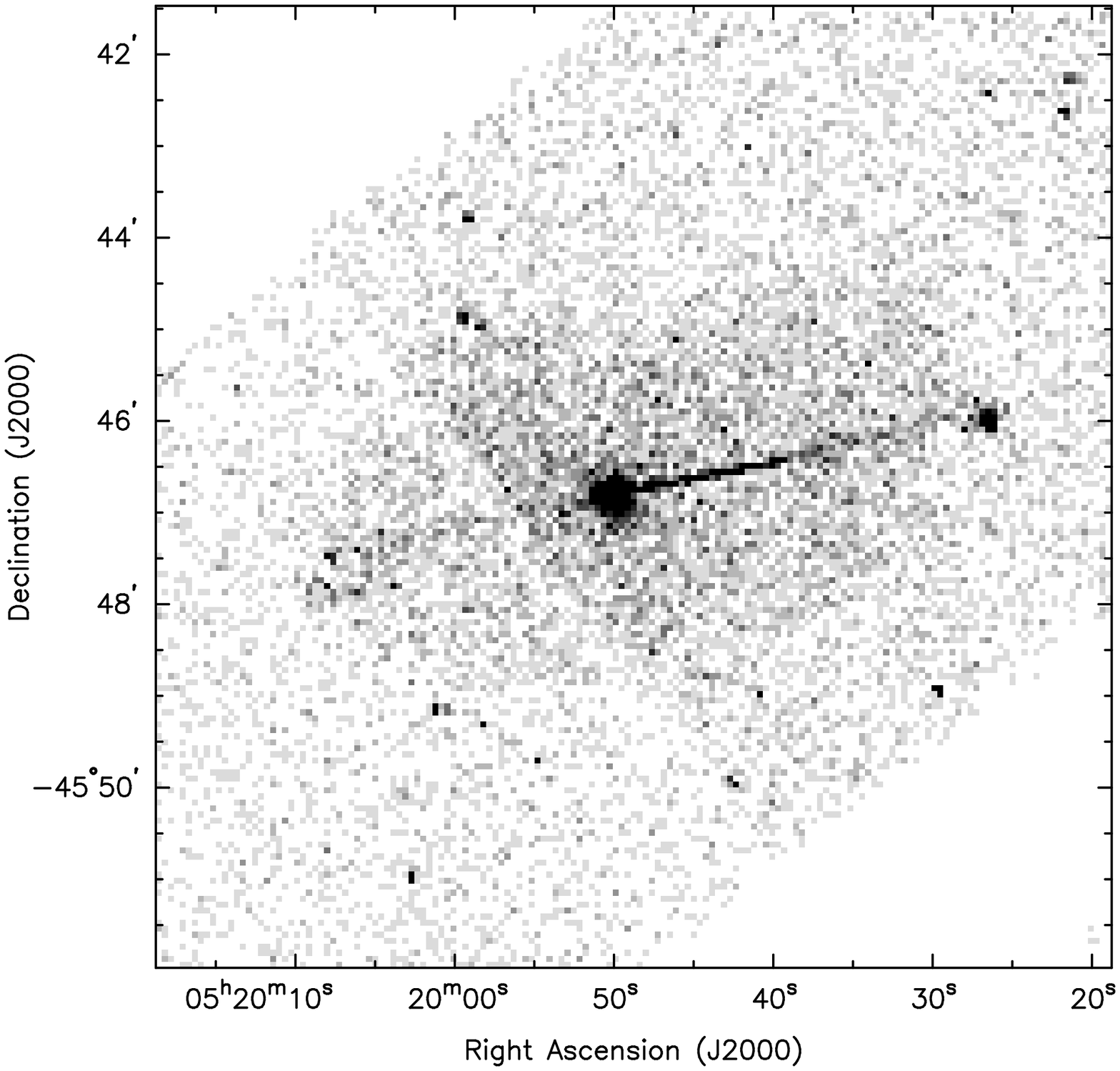}
\hskip 0.5cm
\epsfxsize 7.5cm
\epsfbox{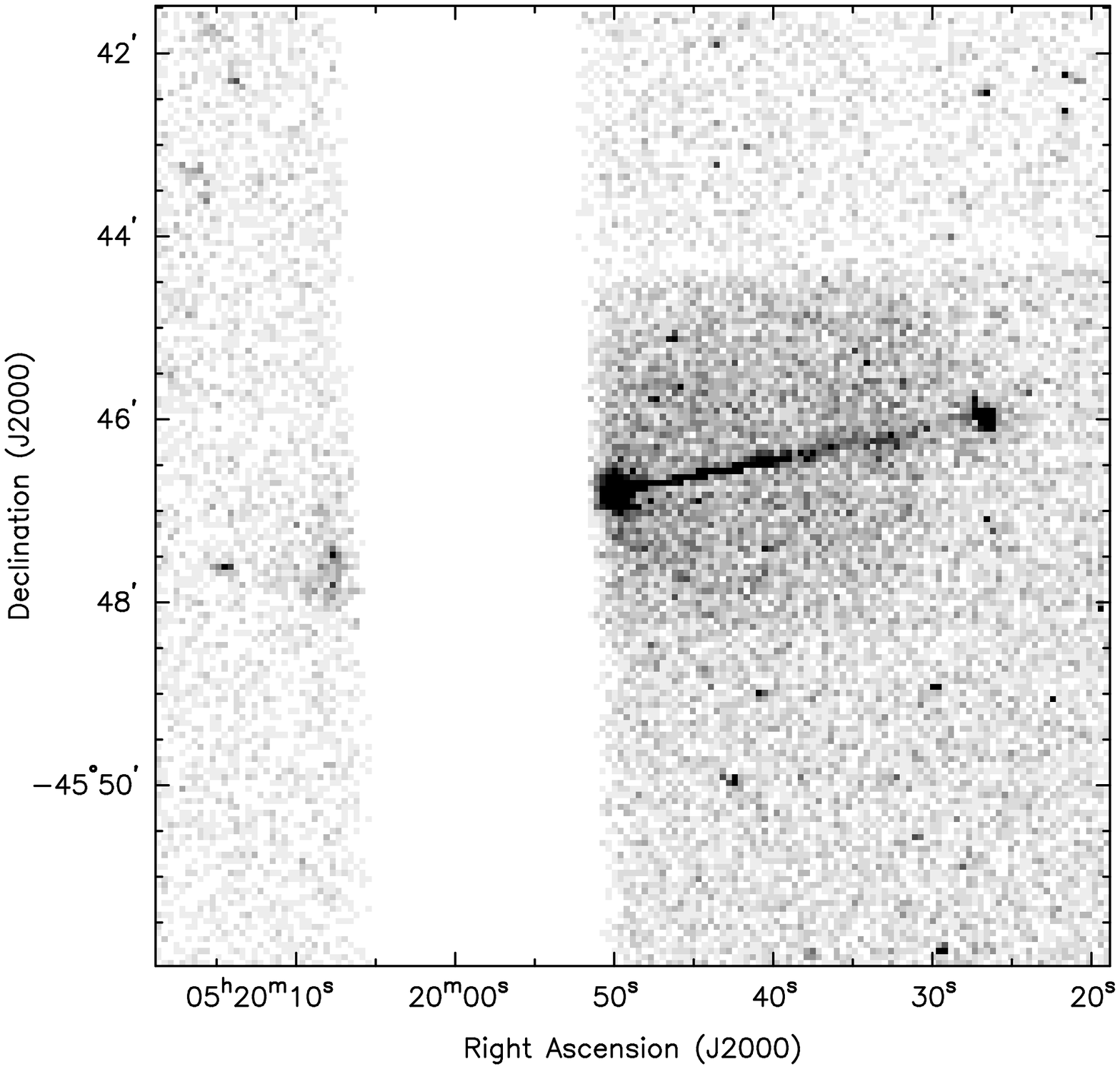}
}
\caption{Raw {\it Chandra} images of Pic A in the 2000 and 2002
  observations in the 0.5--5.0 keV energy range (after filtering and
  removal of the readout streak). Both images are binned by a factor
  8, so that pixels are 3.936 arcsec on a side. In the left-hand image
  the source can be seen to span the ACIS-S2 and S3 chips. In the
  right-hand image the ACIS-I array is to the east and the S array to
  the west: the chip gap between the S3 and S2 chips is visible
  towards the top right of the image. Black is 7 counts pix$^{-1}$ in
  the left-hand image and 14 in the right-hand one. Lobe emission is
  clearly visible in both images.}
\label{rawims}
\end{figure*}

\section{Results and interpretation}

\begin{figure*}
\epsfxsize 16.5cm
\epsfbox{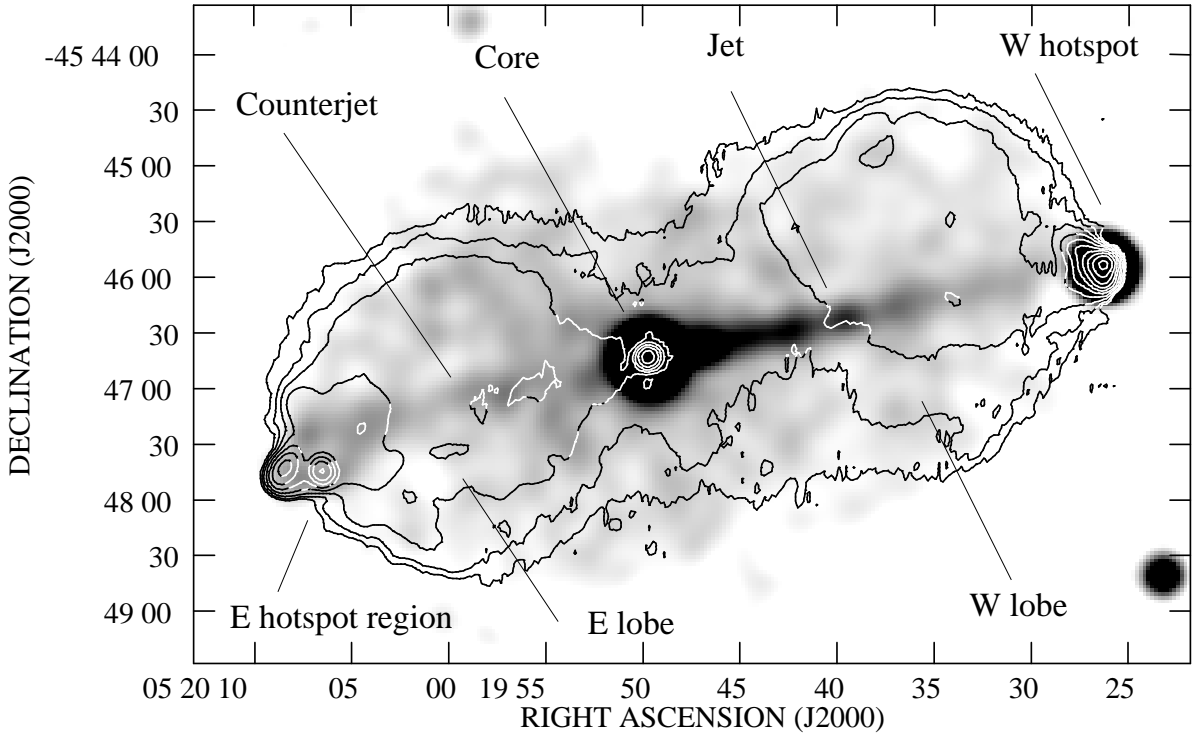}
\caption{X-ray and radio images of Pictor A. The greyscale shows the
  exposure-corrected map made from the 2000 observations in the
  0.5--5.0 keV energy band, smoothed with a Gaussian of FWHM 16
  arcsec, after removal of nearby point sources. The greyscale is
  truncated at the approximate $3\sigma$ level determined from the
  background, $6 \times 10^8$ photons cm$^{-2}$ pixel$^{-1}$; black
  corresponds to $3 \times 10^7$ photons cm$^{-2}$ pixel$^{-1}$.
  Contours are from the 1.5-GHz map of P97 with 7.0-arcsec resolution,
  and are at $10 \times (1, 2, 4\dots)$ mJy beam$^{-1}$. Labels show
  the X-ray components discussed in the text.}
\label{smoothed}
\end{figure*}

Fig.\ \ref{smoothed} shows the main features of the X-ray emission of
Pic A that we wish to discuss in the remainder of the paper: to
illustrate the quality of the spectra we obtain, net counts for the
regions we discuss in the remainder of the section are tabulated in
Table \ref{netcounts}. The brightest compact feature apart from the
core is the E hotspot: this has already been discussed by W01 and will
be discussed in more detail elsewhere (Wilson \etal , in prep.) and we
do not consider it further here. The image in Fig.\ \ref{smoothed},
as well as those of Fig.\ \ref{rawims}, clearly show the jet in the W
lobe, which extends almost all the way to the hotspot, and also hint
at a corresponding linear feature in the E lobe: we will briefly
comment on the nature of the jet and discuss the possible counterjet
in Section \ref{cjet}. On more extended scales, X-ray emission from
the lobes and from a bright region around the E hotspot can clearly be
seen. We discuss the nature of these in the following subsections.

\begin{table}
\caption{Net counts from regions of Pic A}
\begin{center}
\label{netcounts}
\begin{tabular}{lllc}
\hline
Source&Telescope&Epoch&Counts\\
\hline
Core&{\it XMM}&2001 & $(2.9\pm0.02) \times 10^4$\\[3pt]
Jet&{\it Chandra}&2000&$434 \pm 21$\\
&&2002a&$446 \pm 30$\\
&&2002b&$569 \pm 32$\\[3pt]
Counterjet&{\it Chandra}&2000&$19 \pm 9$\\[3pt]
E hotspot&{\it Chandra}&2000&$168 \pm 13$\\[3pt]
E lobe&{\it Chandra}&2000&$975 \pm 49$\\
&{\it XMM} & 2001 & $1715\pm68$\\[3pt]
W lobe&{\it Chandra}&2000&$2471 \pm 193$\\
&&2002a&$2222 \pm 88$\\
&&2002b&$2903 \pm 103$\\
&{\it XMM}& 2001 & $811\pm59$\\
\hline
\end{tabular}
\vskip 8pt
\begin{minipage}{7cm}
Net counts are tabulated in the range used for spectral fits,
i.e. 0.4--7.0 keV for {\it Chandra} and 0.5--5.0 keV for {\it XMM-Newton}.
Dates 2002a and b refer to obsids 3090 and 4369 respectively.
\end{minipage}
\end{center}
\end{table}

\subsection{Integrated lobe properties}

The first test of an inverse-Compton model for the lobes is the flux
density and spectrum of their X-ray emission. Grandi \etal\ have
already shown that the region around the E hotspot has a spectrum
consistent with an IC model. We extracted spectra for
the whole of the W lobe (visible in all three images) and the E lobe
as seen in the 2000 images, where we used two extraction regions with
two different backgrounds to take account of the different responses
and background levels of the S2 and S3 chips. The extraction regions
were ellipses defined to include the emission enclosed by the lowest
contour of the 1.5-GHz radio map shown in Fig.\ \ref{smoothed};
emission from the X-ray nucleus, jet, and hotspot, and several point
sources in the lobes (but not the regions in the E lobe discussed in
Sections \ref{ehs} and \ref{cjet}) was excluded. Spectra and responses
for off-source background regions were also generated. We then fitted
spectra to the data in the energy range 0.4--7.0 keV. Spectral fits
were carried out in two ways. Initially, we used simple background
subtraction within {\sc xspec} 11.3, grouping the spectra so that each
bin contained $>30$ net counts after background subtraction.
Subsequently, we used the ability of {\sc sherpa} to model source and
background
together\footnote{http://asc.harvard.edu/sherpa/threads/sourceandbg/index.html\#indepinstresp},
fitting the backgrounds with an {\it ad hoc} combination of {\it
mekal} models, Gaussians, unabsorbed power laws and heavily absorbed
power laws (the last component being intended to model the turnup in
the background level at high energies). Neither of these two methods
is perfect, as the instrumental background is a combination of
internal/particle background (best dealt with by direct subtraction)
and external, cosmic background (best dealt with by modelling the
background taking into account the differing response). However, the
results derived were almost indistinguishable
within the errors, as shown in Table \ref{loberesults}. In both cases
we fitted three models: a pure power-law model, a pure thermal model
({\it mekal} with 0.5 solar abundance at the redshift of Pic A) and a
combination of the two. In almost all cases the pure thermal model was
the worst fit, and the combined model was the best, though in several
cases the improvement over the pure power-law model was marginal: in
the combined model the thermal component contributes only a small
fraction of the total emission from the lobe, but the power-law index
is made significantly flatter.

For the {\it XMM-Newton} data, we used extraction regions similar in
size to those used for the {\it Chandra} analysis; however, due to the
larger PSF, it was necessary to exclude larger regions for the core,
jet and hotspot. Spectra were extracted for both lobes in the energy
range 0.5--5.0 keV. Our lobe
regions, corresponding to the extent of the low-frequency radio
emission, are much larger than the region studied by Grandi \etal ,
which means that our regions may include a significant contribution
from the PSF wings of the central AGN. We therefore also extracted a
spectrum for the core (using a circle of 2 arcmin) in order to to
estimate its contribution to the lobe spectra. Pileup is not
significant in the core. We fitted the core spectrum with a single
power law, which gave a good fit (reduced $\chi^2 = 1.1$) with $\Gamma
= 1.82\pm0.01$, consistent with the {\it RXTE} results of Eracleous,
Sambruna \& Mushotzky (2000). Our measured flux is a factor of $\sim
2$ below their value, but the AGN is likely to be highly variable. We
then calculated the fraction of the AGN spectrum expected to have been
scattered into the lobe regions. We found that the eastern lobe region
should contain $\sim 3$ per cent of the total AGN emission, and the
western lobe $\sim 2$ per cent, which is a significant fraction of the
flux expected from the lobes based on the {\it Chandra} results. To
account for this AGN component, we included a fixed power law
component in our spectral fits with $\Gamma = 1.82$ and a
normalization obtained by scaling the normalization of the central AGN
appropriately (energy-dependence of the PSF is not a complicating
factor at these radii). In addition to this scattered-AGN component,
we fitted a power-law model to the excess, lobe-related emission. For
the eastern lobe, we obtained a good fit ($\chi^2 = 93$ for 83 d.o.f.)
with a photon index of $1.7\pm0.1$, consistent with the {\it Chandra}
results. For the western lobe, the best fit ($\chi^2 = 42$ for 45
d.o.f.) had a photon index of $2.0\pm0.2$, which is slightly higher
than the {\it Chandra} results, but consistent within the errors with
all but one of the single-component {\it Chandra} fits. We measured
1-keV flux densities of $34.1\pm2.5$ nJy (E) and $19.1\pm2.3$ nJy (W).
As our extraction regions are significantly smaller than the {\it
Chandra} regions due to the pn chip gaps and the larger jet and
hotspot exclusion regions, we scaled these fluxes to account for this
difference, and obtain scaled flux values of $50.6\pm3.8$ nJy (E) and
$45.5\pm5.5$ nJy (W), which are both in good agreement with the {\it
Chandra} results. For both lobes, using a {\it mekal} model resulted
in a marginally worse fit, with temperatures of $\sim 3$ keV,
inconsistent with the best-fitting thermal models from the {\it
Chandra} analysis.

\begin{table*}
\caption{Results of spectral fitting to the lobes}
\label{loberesults}
\begin{tabular}{lrrrrrrrrrrr}
\hline
Dataset/&\multicolumn{3}{c}{Pure power-law}&\multicolumn{3}{c}{Pure \it
  mekal}&\multicolumn{5}{c}{Power-law plus {\it mekal}}\\
Lobe/&Flux&Photon&$\chi^2/n$&Norm.&$kT$&$\chi^2/n$&Flux&Photon&Norm.&$kT$&$\chi^2/n$\\
Method&(nJy)&index&&($\times 10^{-4}$)&(keV)&&(nJy)&index&($\times 10^{-4}$)&(keV)\\
\hline
2000/W/X&$56\pm 2$&$1.75\pm
0.08$&51/46&$3.3_{-0.1}^{+0.2}$&$6.0_{-1.0}^{+1.6}$&67/46&$51\pm
4$&$1.63\pm 0.11$&$0.13 \pm 0.07$&$0.8_{-0.4}^{+0.2}$&47/44\\
2000/W/S&$56 \pm 2$&$1.84\pm 0.10$&114/88&$3.3 \pm
0.2$&$7.0_{-1.5}^{+2.0}$&136/88&$46 \pm 4$&$1.51_{-0.15}^{-0.16}$&$0.3
\pm 0.1$&$0.33_{-0.04}^{+0.08}$&105/86\\
2002/W/X&$54 \pm 1$&$1.77 \pm 0.05$&145/128&$3.0 \pm
0.1$&$4.5_{-0.4}^{+0.7}$&166/128&$47 \pm
2$&$1.64_{-0.04}^{+0.08}$&$0.2 \pm
0.05$&$0.86_{-0.09}^{+0.22}$&130/126\\
All/W/X&$55 \pm 1$&$1.76 \pm 0.04$&196/176&$3.1 \pm
0.1$&$4.9_{-0.5}^{+0.6}$&236/176&$48 \pm 2$&$1.65 \pm 0.05$&$0.18 \pm
0.06$&$0.85_{-0.08}^{+0.16}$&179/174\\
All/W/S&$54 \pm 1$&$1.71 \pm 0.05$&414/359&$3.2 \pm
0.1$&$6.9_{-0.8}^{+1.1}$&456/359&$47_{-2}^{+4}$&$1.49 \pm 0.07$&$0.25\pm0.06$&$0.31_{-0.03}^{+0.04}$&392/357\\
2000/E/X&$55 \pm 2$&$1.57 \pm 0.07$&44/42&$3.4 \pm
0.1$&$7.7_{-1.4}^{+2.3}$&44/42&$49 \pm 3$&$1.44 \pm 0.08$&$0.14 \pm
0.06$&$0.61 \pm 0.17$&37/39\\
2000/E/S&$54 \pm 3$&$1.57 \pm 0.08$&87/75&$3.3 \pm
0.1$&$7.1_{-1.3}^{+2.5}$&88/75&$49_{-4}^{+10}$&$1.48_{-0.11}^{+0.26}$&$0.13
\pm 0.05$&$0.66_{-0.34}^{+0.62}$&83/72\\
\hline
\end{tabular}
\vskip 10pt
\begin{minipage}{18cm}
Fitting statistics and best-fitting parameters are shown. In the first
  column, the date refers to the {\it Chandra} dataset, with the two
  2002 observations being treated as a single dataset. {\sc xspec}
  fits are denoted with an X and {\sc sherpa} fits using background response
  files are labelled with an S. For the {\sc
  sherpa} fits, the fitting statistic includes the fit to the
  background. The errors quoted are the $1\sigma$ errors for one
  interesting parameter. 1-keV flux densities are derived from the
  power-law normalization: the {\it mekal} model normalization is the
  volume-normalized emission measure in standard {\sc xspec} units
  (i.e. $10^{-14}$ cm$^{-5}$): an additional factor of $10^{-4}$ is
  included for convenience of presentation. For the E lobe, two
  regions were used, as described in the text, and the best-fitting
  parameters presented are the results of joint fits, except for the
  fluxes and {\it mekal} normalizations, which are the {\it sums} of
  the normalizations of the two fitted models.
\end{minipage}
\end{table*}

The available data thus strongly favour a non-thermal model. The {\it
Chandra}-derived spectral indices ($\alpha = \Gamma-1$) of the power
laws are in the range 0.5--0.8, which is to first order consistent
with the expectation from an IC model (see below), and also consistent
with the power-law index measured by Grandi \etal\ (2003). The
difference between the flux densities quoted by Grandi \etal\ (2003)
($\sim 17$ nJy: their table 1) and ours ($\sim 54$ nJy for the E lobe)
can be explained entirely in terms of the small region of the E lobe
that Grandi \etal\ considered, since our {\it XMM} analysis shows that
the scaled fluxes are consistent. Our flux density and spectral index
are also consistent with those quoted by Kataoka \& Stawarz (2005) for
the W lobe (derived from the 2000 dataset: Kataoka, priv.\ comm.).

\begin{figure*}
\hbox{
\epsfxsize 7.5cm
\epsfbox{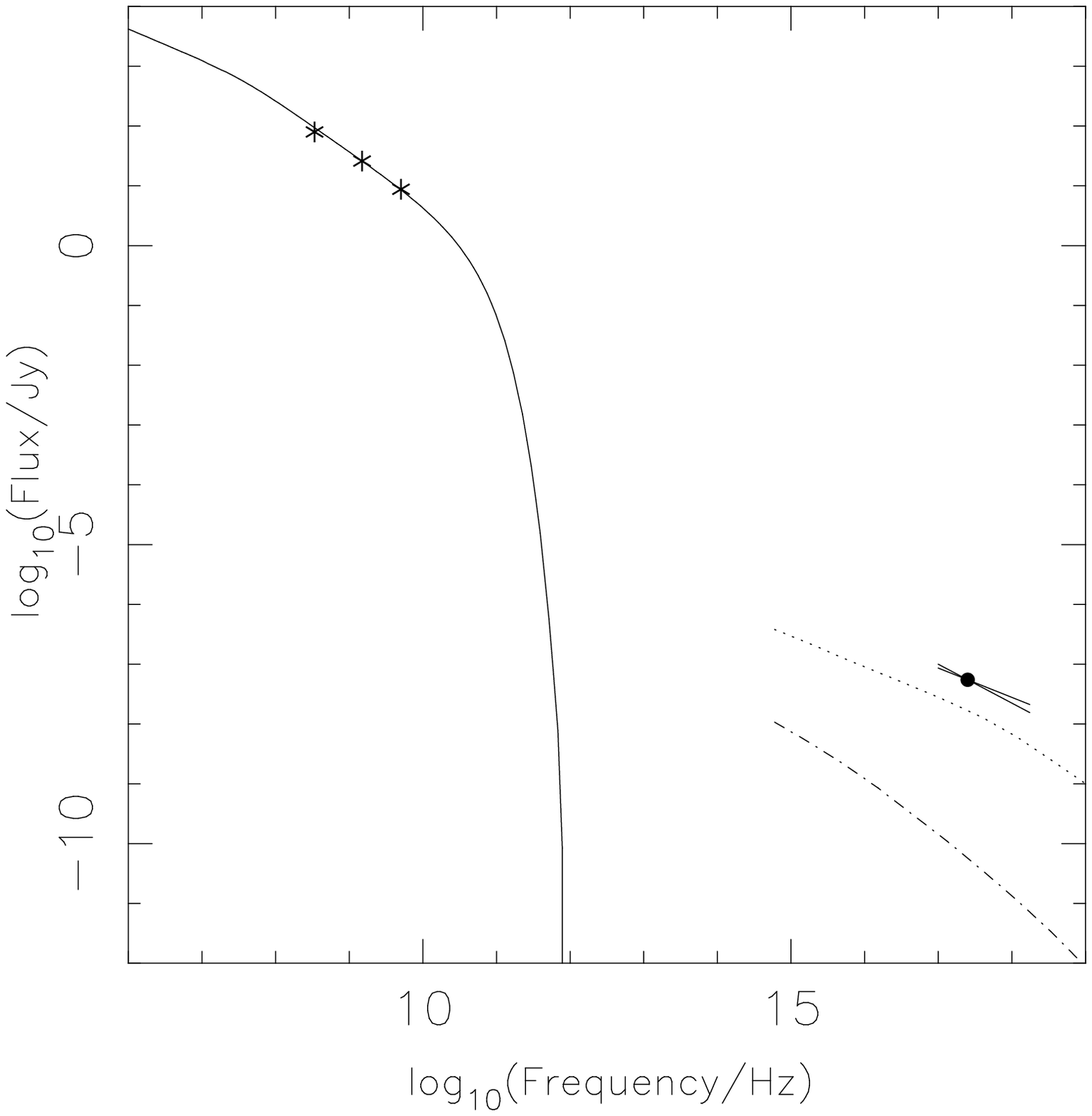}
\epsfxsize 7.5cm
\epsfbox{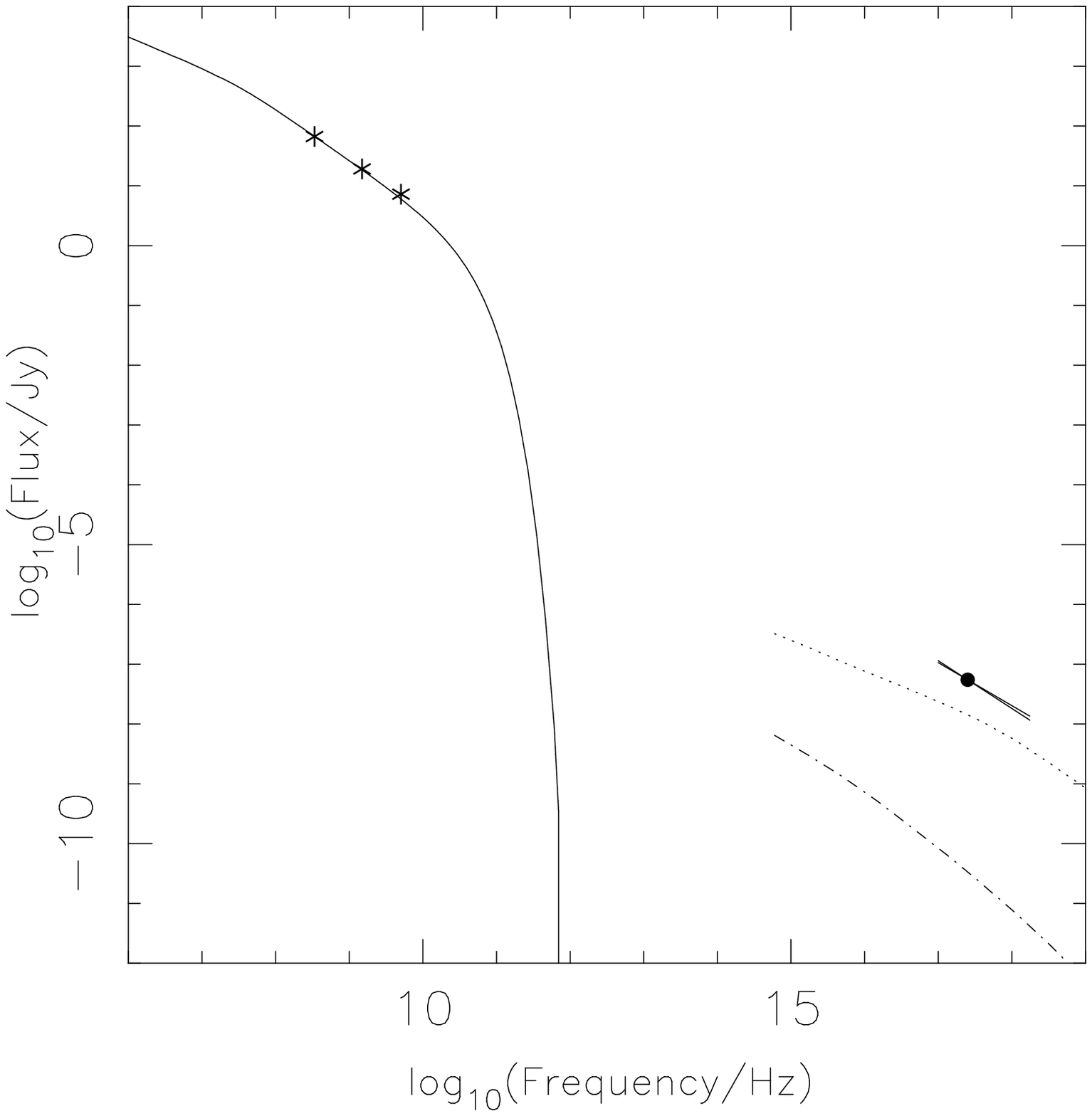}
}
\caption{Broad-band spectra for the lobes of Pic A. The E lobe is on
  the left, the W lobe on the right. The stars show radio data points
  at the three frequencies discussed in the paper. The dot on the
  right of each plot is the X-ray data point at 1 keV and the `bow
  tie' on the dot shows the $1\sigma$ range of the X-ray spectral
  index for the best-fitting power-law model. Error bars are smaller
  than symbols. The solid line is the model synchrotron spectrum used
  (see the text). The dotted line is the predicted flux density for
  inverse-Compton scattering of the CMB as a function of frequency for
  an equipartition magnetic field, and the dot-dashed line shows the
  (negligible) contribution from synchrotron self-Compton emission.}
\label{spectra}
\end{figure*}

We compare the X-ray observations to the predictions of an
inverse-Compton model using the method outlined in Croston \etal\
(2005). In this case, since the integrated spectrum of the source is
well known (P97) we model the electron spectrum to reproduce it in
more detail than was possible in our earlier work: we assume a
low-energy electron energy index of 2, $\gamma_{min} = 10$ as in
Croston \etal, and a break in energy index of 0.7 at $\gamma = 1400$
to give the observed high-frequency synchrotron spectral index $\alpha
= 0.85$. The electron spectrum is normalized using the 1.5-GHz flux
density measured from the maps of P97, using the same extraction
region as was used for the X-ray data. The predicted 1-keV flux
density for an equipartition magnetic field from inverse-Compton
scattering of the CMB (synchrotron self-Compton emission is
negligible) is then 14.5 nJy for the W lobe and 16.8 nJy for the E
lobe. Fig.\ \ref{spectra} shows the broad-band spectra of the lobes,
the model synchrotron spectrum, and the predicted form of the
inverse-Compton emission at equipartition.

Croston \etal\ define a parameter $R$, the ratio between the
observed X-ray flux density at 1 keV and the equipartition prediction:
for the W and E lobes of Pic A $R$ is 3.8 and 3.3 respectively, if we
take the X-ray flux densities derived from the pure power-law fits.
(Note that since the E lobe flux density we quote is a slight
underestimate, the $R$ value is also probably an underestimate.) These
values of $R$, while they lie at the higher end of the distribution
discussed by Croston \etal, are by no means extreme, and in fact lie
close to the median values observed for broad-line objects. The $R$
values would be reduced (to 3.2 and 2.7) if we adopted the flux
densities derived from the power law plus {\it mekal} models. As
discussed by Croston \etal, the true $R$ values for broad-line objects
like Pic A should be lower, because of the effects of projection. Pic
A's axis should be at a relatively small angle to the line of sight,
$\theta \la 45^\circ$, in simple unified models for low-power FRII
objects: Tingay \etal\ (2000) independently estimate $\theta <
51^\circ$ from modelling of the parsec-scale jet, and significant
projection is implied by the observed Laing-Garrington effect (P97).
If Pic A lay at 30$^\circ$ to the line of sight, for example, then the
true $R$ values would be a factor $\sim 1.35$ lower. Without invoking
extreme projection there is no obvious way to reduce $R$ below 2 ($R =
2$ is the median value observed for narrow-line radio galaxies) and so
if the observed X-ray emission is inverse-Compton it requires a
magnetic field strength a factor at least $\sim 1.6$ below the
equipartition value. This is in line with the results of Croston
\etal\ and Kataoka \& Stawarz (2005) and with previous studies of
small numbers of objects (e.g.
Hardcastle \etal\ 2002, Isobe \etal\ 2002, Belsole \etal\ 2004).

The inverse-Compton model predicts that the spectral index measured in
the X-ray should reflect the low-energy electron energy index. In
principle, therefore, the observed values (Table \ref{loberesults})
can tell us something. In practice, the answer depends on the adopted
model. Almost all the fitted power-laws have an energy index less than
the high-frequency radio value (P97) of 0.85, and if
the models containing thermal emission are adopted, the power-law
indices are consistent with values in the range 0.5--0.6. The flatter
X-ray spectral indices are consistent with the tentative observation
by P97 of a low-frequency spectral flattening, and the
lowest values would imply a low-energy electron energy index of
2--2.2, which is consistent with the predictions of simple shock
acceleration models. However, we are reluctant to draw very strong
conclusions from this result, since the inclusion of the thermal
components in the model is by no means required by the fits (and the
discrepant best-fitting temperatures in the models are a cause for
concern). The predicted X-ray spectral index
is also dependent on the position of the low-energy break in the
electron energy index: if the break is at low enough energies, then
the X-ray spectrum will steepen across the {\it Chandra} passband, and
our calculations suggest that this may be the case in Pic A (this can
be seen in Fig.\ \ref{spectra}), though
the position of the low-energy break is not well constrained by the
data. If the X-ray data for the W lobe are fitted with a broken
power-law model (as a first approximation to the true shape of the
inverse-Compton prediction), with the low-energy spectral index fixed
to 0.5, then it is interesting, though of course not conclusive, that
the best-fitting high-energy spectral index is $0.84\pm 0.06$, in
excellent agreement with the observed high-frequency radio spectral
index for the source.

Finally, it is worth noting the difference between the fitted X-ray
spectral indices for the two lobes. The E lobe is flatter in all fits
than the W lobe. This is in the opposite sense to the trend seen in
the radio at higher frequencies (P97). We comment on
this further below.

\subsection{Lobe properties as a function of position}

\begin{figure}
\epsfxsize 8.5cm
\epsfbox{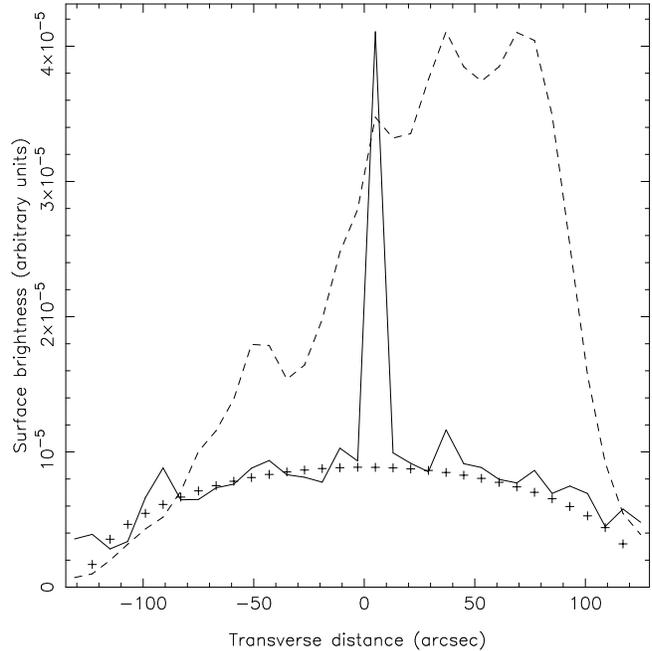}
\caption{The surface brightness profile across the W lobe of Pic A.
  The profile is extracted from a region 45 arcsec wide in the
  broadest part of the lobe, in a direction perpendicular to that of
  the jet, and binned in 8-arcsec bins. The solid line shows the
  profile taken from the unsmoothed exposure-corrected X-ray data: the
  central peak is the jet. The dashed line is the radio emission from
  the 7.5-arcsec resolution 1.5-GHz map; the jet is visible, as is the
  region of high surface brightness in the NW part of the lobe
  (positive values on the $x$ axis). Crosses show the expected
  emission from a uniform-emissivity projected sphere or cylinder with a radius
  of 120 arcsec centred at 0 arcsec. Although the normalization has
  been adjusted to reflect the normalization of the X-ray data, these
  points are in no sense a fit to the data and are plotted simply to
  guide the eye.}
\label{slice}
\end{figure}

Fig.\ \ref{smoothed} shows that the surface brightness of the lobes in
the X-ray is very much more uniform than in the radio. As an example
of the surface brightness profile across the lobe, we show a slice
taken from the W lobe between 80 and 125 arcsec from the core (i.e.
roughly the broadest part of the lobe) in Fig.\
\ref{slice}. Large variations in the radio surface brightness
are not reflected in the X-ray data: the X-ray profile across the lobe
is well modelled as a projection of a sphere or cylinder, while this
is not at all the case in the radio.

\begin{figure*}
\epsfxsize 16.5cm
\epsfbox{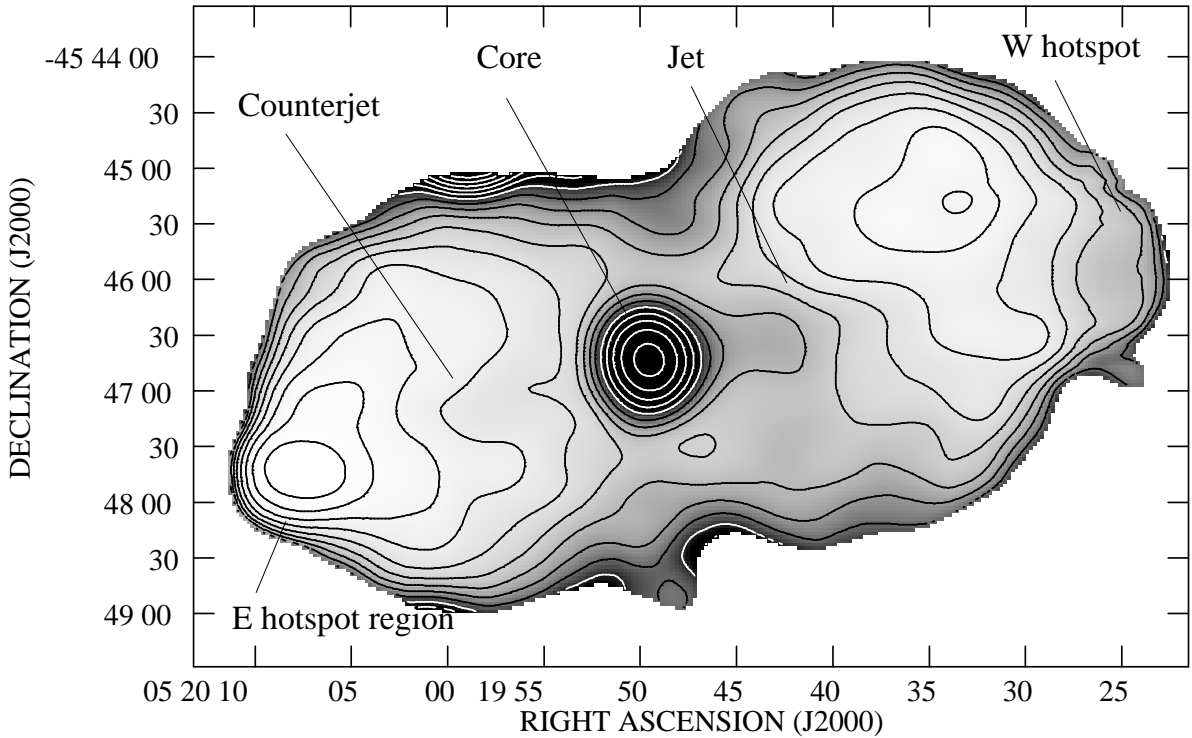}
\caption{The ratio between the exposure-corrected X-ray in the 0.5--5.0
  keV energy band and the 330-MHz radio emission of Pictor A.
  Greyscale and contours both show the ratio (arbitrary units). The
  greyscale is plotted in the sense that the lighted areas are
  relatively brighter in the radio and the darkest relatively brighter
  in the X-ray. Contours increase logarithmically by a factor
  $\sqrt{2}$. No data below the $5\sigma$ level of the
  330-MHz map are shown. Labels indicate the positions of features
  discussed in the text.}
\label{ratio}
\end{figure*}

The high signal-to-noise of the X-ray data allows us to make a map of
the X-ray/radio ratio. To do this we used the $30\times 30$-arcsec
resolution 330-MHz radio map of P97, as being the map (at useful
angular resolution) that best traces the low-energy electrons, and
convolved the exposure-corrected X-ray map from the 2000 {\it Chandra}
observation to the same resolution. The ratio of these two maps is
shown in Fig.\ \ref{ratio} (very similar results are obtained using
the 2002 data, but of course these show the W lobe region only). A
number of interesting features can be seen here:

\begin{enumerate}
\item The X-ray jet and W hotspot are clearly distinguished from the
  other regions of emission, as expected.
\item The `counterjet' region can also be seen (best seen as an
  indentation in the contours in the E lobe).
\item Although the E hotspot region is visible as a high-surface
  brightness region in Fig.\ \ref{smoothed}, its X-ray/radio ratio
  is actually lower than in the lobes as a whole.
\item The X-ray/radio ratio in the lobes is
  higher (by a factor up to $\sim 3$) close to the
  nucleus relative to the value in the centre of the lobes.
\item The X-ray/radio ratio is also higher at the very edges of the lobes.
\end{enumerate}

We return to the jets and hotspots in later sections, but here we will
focus on the explanation for the positional differences in X-ray/radio
ratio in the lobes, which are reminiscent of the results obtained by
Isobe \etal\ (2002) for 3C\,452. For a constant magnetic field
strength throughout the lobe, and a constant ratio between the number
densities of the electron populations responsible for the synchrotron
radiation and those responsible for the inverse-Compton emission, the
X-ray/radio ratio should be constant across the lobe. There are thus
at least three possible models for the observed spatial
dependence of the ratio, which are not mutually exclusive:

\begin{enumerate}
\item Some other emission process boosts the X-ray emission in the
  inner regions.
\item The radio flux measurements
under-represent the low-energy electrons in the central parts of the
source relative to the outer parts.
\item The
magnetic field strength relative to the equipartition value is varying
as a function of position, such that the inner lobes are more
electron-dominated.
\end{enumerate}

We now discuss each of these models in turn.

\subsubsection{Additional emission components}

Model (i) requires a plausible additional emission process. Two
such processes are thermal emission and inverse-Compton scattering of
nuclear photons (e.g. Brunetti \etal\ 1997). To test these models we
need to see whether there are differences between the X-ray spectra of
the inner and outer lobe regions. We extracted two matched spectral
regions (circles of 50-arcsec radius) centred on the regions of high
and low X-ray/radio ratio in the E lobe from the 2002 {\it Chandra}
data. Each circular region represented slightly under a quarter of the
total flux from the lobe. We fitted pure power-law models and power
law plus {\it mekal} models to both datasets; as expected, both were
good fits. The best-fitting power-law indices were different, but not
significantly so ($\alpha = 0.8 \pm 0.1$ for the region close to the
core compared to $0.9 \pm 0.1$ for the region further out in the lobe)
and although the best-fitting thermal component in the combined model
(which gave flatter spectral indices, as expected) had a higher
normalization in the inner extraction region, the power-law emission
was still dominant, accounting for $>90$ per cent of the flux. In the
W lobe, as seen with the 2000 data, the situation is somewhat more
extreme: examining separately the regions that lie on the S3 and S2
chips (which happen to correspond roughly to the regions of higher and
lower X-ray to radio ratio) the best-fitting photon indices are
respectively $0.4\pm0.1$ and $0.7\pm0.1$, and a {\it mekal} model with
plausible temperature in the S3 (central) region has negligible
normalization.

The fits with thermal models imply that we can certainly rule out the
model in which thermal emission dominates in the inner regions and
accounts for the extra X-ray emission. A nuclear inverse-Compton model
remains possible based on these spectral constraints, and the flatter
spectra in the inner regions would be compatible with this if the
low-energy electron spectrum flattens, since the electrons responsible
for the scattering of nuclear photons have lower energies (and so
presumably a flatter electron energy spectrum) than those
scattering the CMB photons. However, for the nuclear photons to be
comparable in energy density to the CMB at the distances we observe
(up to around 80 arcsec in projection for the E lobe, or $>60$ kpc)
the nuclear luminosity in far-IR photons would have to be $>10^{40}$
W, which seems high for Pic A (comparable to the bolometric luminosity
of a luminous quasar). Moreover, we would expect a strong side-to-side
asymmetry, in the sense that the counterjet-side lobe would be
significantly (by a factor $\sim 7$ or more, assuming $\theta <
45^\circ$) more dominated by nuclear emission, if the source is at the
angles to the line of sight expected from unified models and the
jet/counterjet asymmetry is due to beaming: in fact, if anything, the
jet-side lobe has a larger region of high X-ray/radio ratio.
Accordingly, we consider it unlikely that nuclear inverse-Compton
emission is very important in this source, though it may make a
contribution to the region of flatter X-ray spectrum in the inner part
of the W lobe, and thus to the overall flatter X-ray spectrum of the W
lobe discussed above.

\subsubsection{Electron spectrum variation}

If model (i) can be dismissed, we can consider
explanations in which the X-ray emission is dominated by
inverse-Compton scattering of the CMB, and is giving us information
about a single low-energy electron population. Model (ii), in
which it is positional differences in the electron spectrum that are responsible
for the differences in X-ray/radio ratio, is favoured at first sight
by the fact that the radio spectrum of the inner part of the lobes is
steeper than the outer part even at the lowest frequencies (P97): in
fact, the relatively X-ray bright inner parts and edges correlate
extremely well with the steeper-spectrum ($\alpha^{0.33}_{1.5} \sim
0.9$ rather than $0.75$) regions of the lobe shown in fig.\ 5 of P97.
However, it is hard to explain the {\it magnitude} of the difference
between the different regions of the lobes purely in terms of the
changes in the electron spectrum implied by the differences in radio
spectral index. Emission at 330 MHz traces electrons around $\gamma =
6000$, if the magnetic field implied by the inverse-Compton model for
the whole lobe is adopted, while the inverse-Compton emission traces
electrons around $\gamma = 1000$: thus, since critical frequency for
synchrotron emission goes as $\gamma^2$, changes in X-ray/radio ratio
of a factor 3 would require changes in the synchrotron spectral index
between 10 MHz and 330 MHz across the lobe of
$\log(3)/\log(6000^2/1000^2) \approx 0.3$, if magnetic field strength
is held constant, which is greater than the observed change at higher
frequencies. Even if we allow the magnetic field strength to take the
equipartition value (taking into account the fact that a steeper
spectrum also implies more low-energy electrons) a calculation using
the inverse-Compton code produces similar conclusions -- changes in
the low-frequency spectral index of the magnitude observed in the
low-frequency maps of P97 ought to produce at most a factor 2
difference in the X-ray-to-radio ratio for 330-MHz radio emission,
which is somewhat less than what is observed. Only if there are more
low-energy electrons than a simple extrapolation from higher frequency
would predict can we explain the observed differences purely in terms
of numbers of low-energy electrons in a homogenous model. However,
this calculation of course assumes a single spectral index (and
magnetic field strength) along the line of sight through the lobe
region of interest, which is simplistic: the measured spectral index
of the lowest-$\alpha$ parts of the lobe is almost certainly an
underestimate of the true lowest $\alpha$ present, and so it seems
plausible that more of the observed difference could be accounted for
with a more detailed source model. A version of explanation (ii) may
still be the correct one, even though we have not been able to
contrive one that gives quantitatively correct answers.

\subsubsection{Magnetic field variation}

There is no {\it a priori} reason to reject explanation (iii), in
which magnetic field varies as a function of position, and only modest
changes in the ratio of actual to equipartition magnetic fields are
needed to produce the observed differences -- around a factor 1.5, if
combined with the electron energy spectrum changes discussed above.
This would imply that the magnetic field strengths overall are closer
to equipartition in the parts of the lobes nearest the hotspots and
fall below the equipartition values in the more distant (older?) parts
of the lobe. The lower magnetic field reduces the synchrotron
emissivity and so increases the X-ray/radio ratio; the low-energy
electron population is essentially the same throughout the lobe,
accounting for the relatively uniform inverse-Compton surface
brightness (e.g.\ Fig.\ \ref{slice}). Note that the X-ray
observations, in this picture, are inconsistent with the models
discussed by P97 in which the magnetic field strengths are {\it
higher} at the periphery of the lobes.

It is interesting to ask whether explanation (iii) can also explain the
correlation between high X-ray/radio ratio and steep radio spectrum
without any need for electron population variations: is it
possible that there is a single electron population throughout the
lobe, and the changes in both radio surface brightness and spectral
index are a result of varying magnetic field strength? Such a model is
motivated by the results of Katz-Stone, Rudnick \& Anderson (1993),
which show that it is possible that a single electron energy spectrum
describes all spectral regions of Cygnus A. We can test this model in
Pic A if we assume (unrealistically) that the magnetic field along a
given line of sight is constant: then (e.g. Longair 1994 eq. 18.50) if
the electron energy spectrum at low energies is a power law with index
$p$, the magnetic field strength $B$ is proportional to the X-ray/radio
ratio to the power $-2/(p+1)$. (The number density of electrons along
the line of sight has been cancelled by dividing through by the X-ray
surface brightness.) The observed changes in X-ray/radio ratio of a
factor 3 thus should correspond to magnetic field changes of at most a
factor 2 (depending on the choice of $p$, but assuming $p>2$) so that,
as critical frequency for a given electron energy goes as $B^{-1}$,
the low-frequency ($\alpha^{0.33}_{1.5}$) spectral indices we observe
in the regions of highest X-ray/radio ratio would correspond to
$\alpha^{0.66}_{2.8}$ in the regions of lowest X-ray/radio ratio. As
we know from the maps of P97 that
$\alpha^{1.5}_{4.8}$ in the high-surface-brightness regions of the
lobe is $\sim 0.8$, this model does not seem to be viable: we need
larger variations in the magnetic field strength than are implied by
the range of X-ray to radio surface brightness to shift steep-spectrum
material into the low-frequency band.

\begin{figure*}
\hbox{
\epsfysize 9cm
\epsfbox{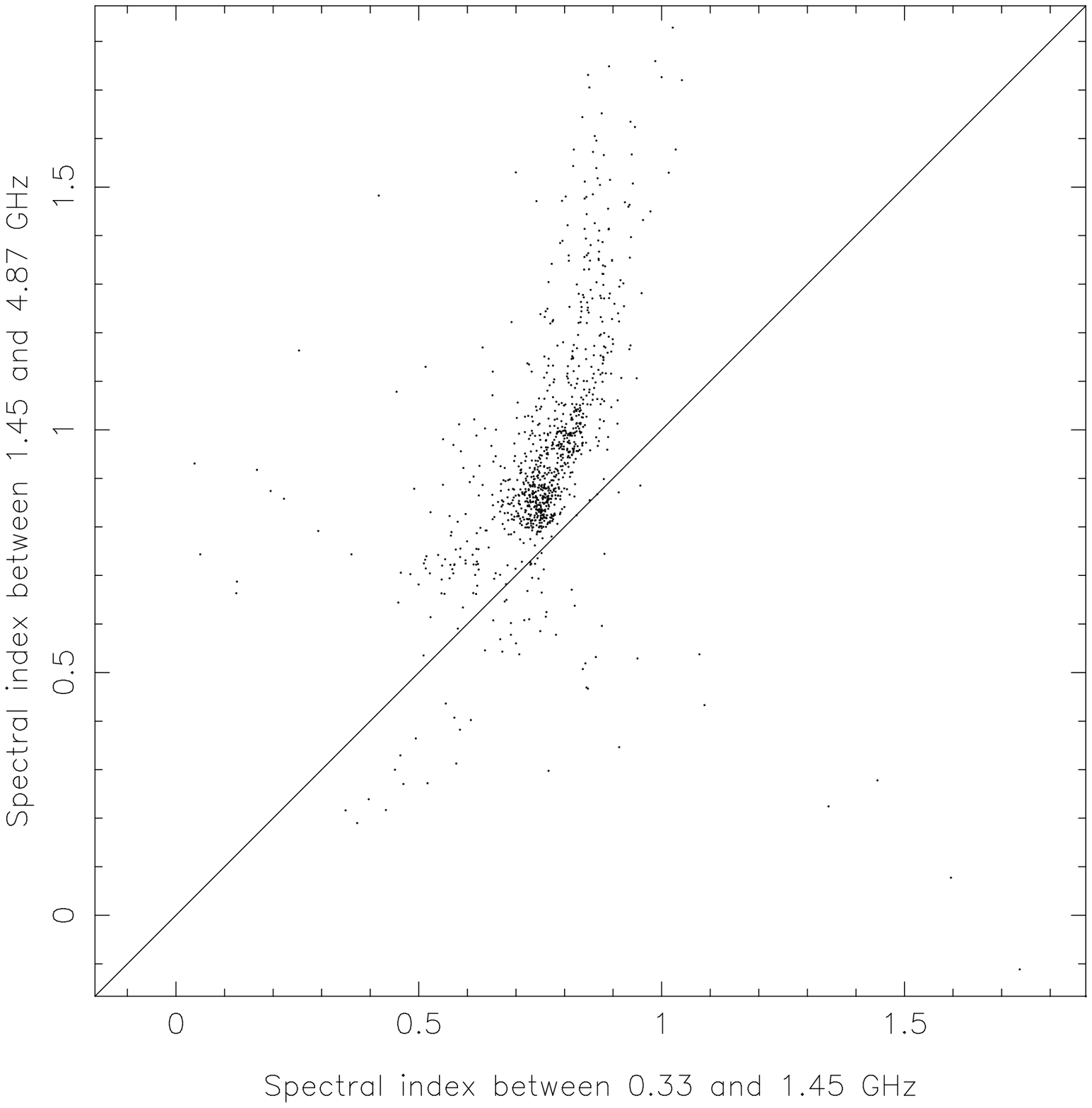}
\hskip 10pt \epsfysize 9cm
\epsfbox{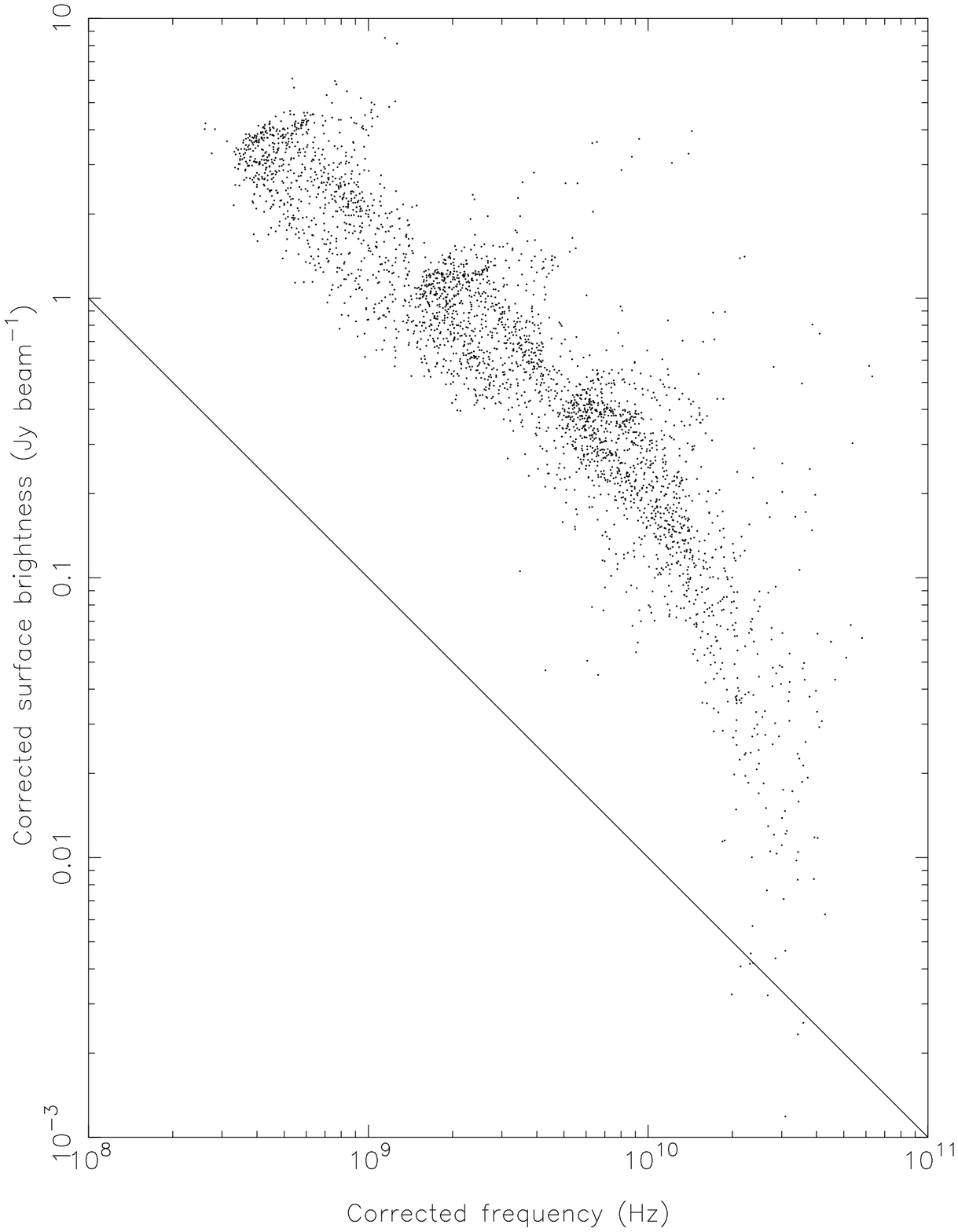}
}
\caption{Colour-colour plots and `corrected' spectrum for Pic A. Left:
  a colour-colour plot of the type described by Katz-Stone \etal\
  (1993) for the 330-MHz, 1.5-GHz and 4.8-GHz Pic A data of P97. The
  maps used have a resolution of 30 arcsec and the
  plot shows the spectral indices for each $10\times 10$ arcsec pixel
  in which the signal exceeds the off-source $3\sigma$ level in each
  map. The line shows the locus of power-law spectra: the points that
  lie below it (i.e. high-frequency spectrum is flatter than
  low-frequency) are mostly associated with the core of Pic A. Right:
  the spectrum of the whole source after `correction' of the surface
  brightness and frequency of each point at each frequency to the
  value expected for the magnetic field strength at the brightest part
  of the lobe, using the magnetic field strength inferred from the
  X-ray/radio ratio as described in the text. For clarity this plot
  does not show points in the core or hotspot regions. If there were a
  single electron spectrum, we would expect these points to lie on or
  near a single (probably curved) line: in fact we see that the
  low-surface-brightness points at 330 MHz and 1.5 GHz lie below the
  expectation for such a model. The solid line shows a power law with
  $\alpha=1.0$ to facilitate by-eye estimation of the slope at various
  frequencies.}
\label{cc}
\end{figure*}

Another way of visualising the
problem is provided by point-to-point radio spectral analysis of the
type carried out by Katz-Stone \etal , and we show this in Fig.
\ref{cc}: although the colour-colour plot is consistent with the
findings of Katz-Stone \etal\ for Cygnus A, an attempt to `correct'
each point using the X-ray/radio ratio to the expected spectrum for a
fixed magnetic field strength does not produce a uniform spectrum.

\subsubsection{Summary of lobe models}

We find no evidence for additional emission components in the high
X-ray/radio ratio regions of Pic A (model (i)) and so consider models
in which Fig.\ \ref{ratio} is telling us about real differences in
lobe inverse-Compton emissivity. Our discussion of model (iii) shows
that magnetic field strength changes can easily account for the
X-ray/radio differences, but that there is genuine variation in the
point-to-point properties of the high-energy end of the electron
spectrum in Pic A; on the other hand, our discussion of model (ii)
suggests that the electron energy spectrum differences cannot in
themselves account for the observations if the spectrum is
conventionally shaped (flatter at lower energies). Our favoured model
is thus a combination of (ii) and (iii): both the electron energy
spectrum and the magnetic field vary as a function of position in the
lobe, with the magnetic field being somewhat weaker, and the electron
energy spectrum more depleted of high-energy electrons, as a function
of distance from the hotspot. Such a model can explain both the
observed changes in spectral index and X-ray/radio ratio.

Finally we emphasise that the diagnostics we have used, which make use
of two-dimensional projected quantities, are probably a crude
representation of the true three-dimensional variation of physical
conditions inside the source. Fully modelling Pic A in three
dimensions would involve many assumptions, given the lack of symmetry
in the source, and we have not attempted to do so. We have, however, made
simple three-dimensional models treating the source as a uniformly
filled sphere of electrons (this seems reasonable given the results
shown in Fig.\ \ref{slice}) with radial variation in magnetic field
and using realistic electron energy spectra. This allows us to discard
the assumption of a single magnetic field strength along the line of
sight. These models give very similar results to our conclusion above
with respect to model (iii): models that can reproduce the magnitude of
the X-ray/radio ratio variation across the source purely in terms of a
variation of the magnetic field strength also produce a much greater
range of spectral indices (and in particular much steeper
high-frequency spectral indices) than are observed in the data.

\subsection{Lobe dynamics and depolarization}

If we adopt the magnetic field strengths determined above for the
whole lobe, then the internal pressure in the lobes is of the order $2
\times 10^{-13}$ Pa (somewhat less if projection is important). The
pressure of any external environment must be less than or equal to
this value unless there is a significant contribution to the internal
energy density from other sources, such as protons. This implies a
local environment comparable to, at most, a moderately rich group of
galaxies, a conclusion consistent with the low temperatures found in
the joint power-law/{\it mekal} fits to the lobes and with the results
of Miller \etal\ (1999). We do not find either the {\it Chandra} nor
{\it XMM-Newton} data to show any convincing evidence for a spatial
detection of thermal emission on the scales of the lobes: this is not
surprising, as the surface brightness of such emission would be low.
Grandi \etal\ (2003) suggest that there may be some extended thermal
emission in the central arcmin of the source, but we see no strong
evidence for this in the {\it Chandra} data.

Can a poor thermal environment such as this provide the increased
dispersion in the rotation measure (RM) on the counterjet side
required to produce the observed low-frequency depolarization of the
counterjet lobe? This question may be ill-posed -- P97 suggest that
the excess RM dispersion they observe is {\it not} enough to produce
all the depolarization, so that the Faraday screen may not be fully
resolved by their observations. Setting that aside, we can make a very
crude estimate of the required properties of the medium by assuming
that the density of the medium along the additional path between the
jet and counterjet-side lobes is such as to give $p = 2 \times
10^{-13}$ Pa for $kT \sim 0.5$ keV, i.e. $n \sim 2.5 \times 10^3$
m$^{-3}$. If we take the scale of the RM variations to be $\sim 10$
arcsec (P97), or 7 kpc, which is similar to what is observed in other
sources, then the projected magnetic field strength in the thermal
environment required to produce the excess RM dispersion is of the
order 0.04 nT, corresponding to a magnetic field energy density
significantly less than the thermal energy density. Thus there is no
difficulty in principle in producing the observed depolarization
properties if the lobe is close to pressure balance with the external
medium: the magnetic field energy density would start to be comparable
to the thermal energy density, making such a model difficult to
sustain, only if the external thermal density were about an order of
magnitude lower than our estimate.

\subsection{The E hotspot region}
\label{ehs}

Although, as discussed above, the X-ray/radio ratio for the E hotspot
is lower than elsewhere in the source, this does not necessarily imply
a low $R$ value. To investigate the nature of the E hotspot region in
more detail we extracted a spectrum from the 2000 dataset from the
region corresponding to the region of higher radio and X-ray surface
brightness seen in Fig. \ref{smoothed}, using an elliptical extraction
region with major axis 40 arcsec and minor axis 23 arcsec, and taking
background from an adjacent identical region in the lobe. We fitted a
power-law model to the region, obtaining a spectral index of $0.7 \pm
0.3$ and a total flux density of $6 \pm 1$ nJy. As shown in the
spectral index maps of P97, this region is relatively
flat-spectrum, and we modelled the radio data at 330 MHz, 1.5 and 4.8
GHz with a standard continuous injection model (Heavens \&
Meisenheimer 1987) with a low-energy electron energy index of 2
($\alpha = 0.5$) which provides a good fit to the data with an
electron energy break around $\gamma = 6000$, using an ellipsoidal
volume for the region. In this model, the
expected inverse-Compton flux density from the region at equipartition
is 0.64 nJy, so that the $R$ value for the hotspot is around 9, much
higher than that for the lobes as a whole. In terms of the X-ray/radio
maps of Fig.\ \ref{ratio}, this illustrates the importance of taking
into account the geometry of the region when converting between ratio
and $R$ value.

\begin{figure*}
\epsfxsize 12cm
\epsfbox{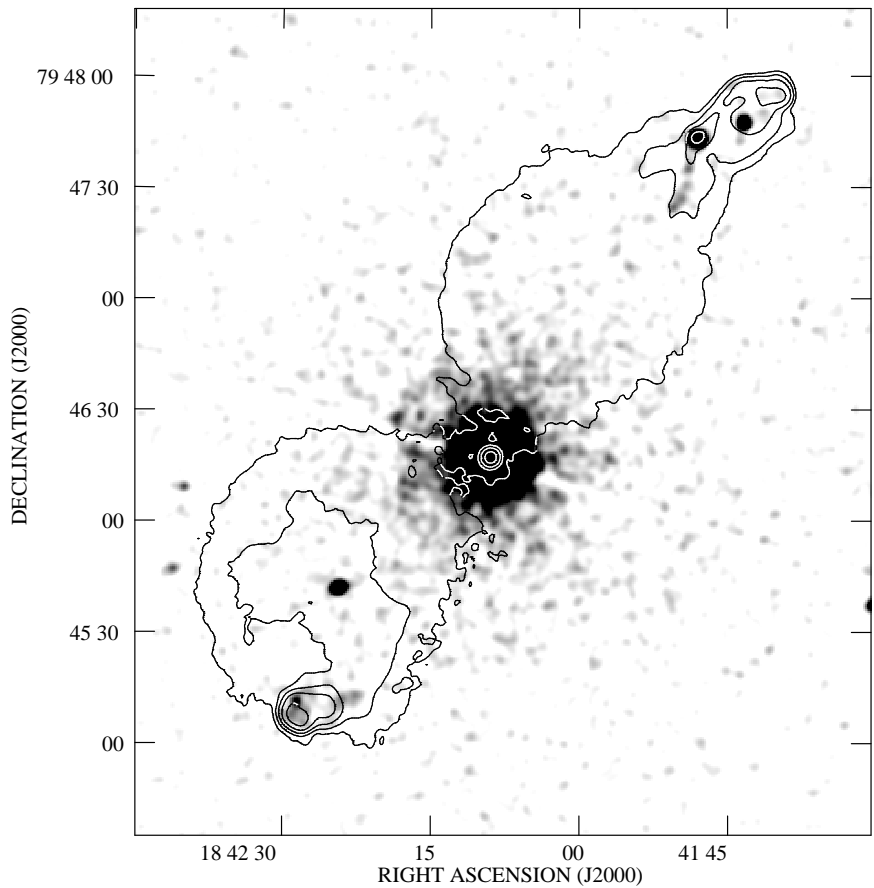}
\caption{Diffuse X-ray emission from the S hotspot region of the
  $z=0.0569$ broad-line radio galaxy 3C\,390.3. Greyscale shows a
  27-ks filtered {\it Chandra} ACIS-S3 image in the energy range 0.5--5.0
  keV, smoothed with a Gaussian of FWHM 4.7 arcsec, with black being
  0.5 counts per 1-arcsec pixel: superposed are
  contours from a 2.8-arcsec resolution 1.5-GHz radio map (Leahy \&
  Perley 1991) at $1.5 \times (1,4,16\dots)$ mJy beam$^{-1}$.
  3C\,390.3 shows X-ray emission from both N and S hotspot regions
  (Hardcastle \etal\ 2004) but while the northern X-ray emission comes
  predominantly from compact features in the N hotspot and jet, the
  X-ray counterpart of the S hotspot is associated with the general
  brighter region around the hotspot, as in Pic A.}
\label{390.3}
\end{figure*}

There are two possible explanations for the high $R$ value in the E
hotspot region: either the emission process is inverse-Compton, but
the magnetic field strength is lower relative to the equipartition
value than in the rest of the lobe, or else a second emission process
is operating in the region around the hotspot. There is no evidence
for thermal emission in the spectrum ({\it mekal} models give a high
temperature and a poorer fit), but the poorly constrained spectral
index is consistent with either an inverse-Compton or a synchrotron
model. An interesting question is therefore whether synchrotron
emission from the hotspot region can have an effect. To produce the
observed $R$ value, the synchrotron emission would have to be about
twice as bright as inverse-Compton from the same region. Synchrotron
emission probably dominates the W hotspot (W01) but the X-ray emission
from the E hotspot is not associated with any of the compact
components of radio emission seen in the images of P97, whereas there
is very good agreement between the bright, compact radio and X-ray
structures in the W hotspot. The spectral index in the hotspot region
between 5-GHz radio and X-ray is 1.1, which is entirely consistent
with a synchrotron model. If the X-ray emission in the E hotspot
were synchrotron in origin, there would have to be distributed
particle acceleration over a 50-kpc region without obvious radio
counterparts for particle acceleration sites: however, distributed
particle acceleration on comparable scales is possible in the jets of
low-power FRI sources (e.g. Evans \etal\ 2005) and even in the W
hotspot a simple model involving particle acceleration at a single
point is inconsistent with observations of the optical filament (P97).
At the time of writing we know of two other sources that show
anomalously bright diffuse X-ray emission, extended over several kpc
and not primarily associated with compact radio features, near one
hotspot complex: the W hotspot region of 3C\,403 (Kraft \etal\ 2005)
and the S hotspots of 3C\,390.3 (Hardcastle \etal\ 2004: Fig.\
\ref{390.3}). As these two are both, like Pic A, low-power FRII
sources with bright, almost certainly synchrotron X-ray hotspots and
jets in their other lobes, the association between synchrotron
emission and diffuse hotspot-associated emission of the type seen in
Pic A East seems strong, and leads us to believe that the diffuse
emission is most likely also synchrotron in origin. If the conclusions
of Hardcastle \etal\ (2004) are correct, this sort of diffuse emission
would be expected to be seen only in low-luminosity sources like Pic
A. It is not at all clear why this type of feature should be seen only
on the {\it counterjet} side of the sources in which it is detected,
but only a few FRII sources of comparable luminosity have been
observed at the time of writing, and it may be that this apparent
trend (which would be difficult to explain in the context of unified
models in which one-sided jets are an indication of beaming) will
disappear with better statistics. One possible explanation is that
modestly relativistic speeds persist in the backflow beyond the
compact hotspots: Doppler boosting effects would then suppress X-ray
emission on the jet side and enhance it on the counterjet side. A
quantitative test of this idea must await the detection of more
low-luminosity X-ray hotspots.

\subsection{The counterjet}
\label{cjet}

W01 discuss synchrotron and beamed inverse-Compton models for the
bright jet seen in their images. We will not discuss the jet as seen
in the 2002 data in great detail here, as it will be the subject of
other work (Wilson \etal\ in prep.). Here we note only that the X-ray
spectrum of the jet in the 2002 data is well-constrained and steep. We
extract a spectrum from the narrow region of the jet, within 2 arcmin
of the nucleus, using adjacent matching rectangular background
regions, and find that the best-fitting power-law model (with fixed
Galactic absorption, as in our other fits) has a photon index of $1.97
\pm 0.07$. The 1-keV flux density in this extraction region is $11.8
\pm 0.4$ nJy. This steep spectrum, if we accept an inverse-Compton
model for the lobes, makes a beamed inverse-Compton model for the jet
very hard to accept, as it requires the low-energy electron population
in the jet to have an energy index significantly steeper than that in
the lobes. On the other hand, the radio and X-ray properties of the
jet region can be well described by a standard continuous-injection
synchrotron model, with a spectral break at a few $\times 10^{11}$ Hz.
Other factors that seem to us to strongly favour a synchrotron model
are the knotty nature of the X-ray jet, which cannot be well explained
in an inverse-Compton model (e.g. Tavecchio, Ghisellini \& Celotti
2003) and the fact that its X-ray surface brightness drops markedly as
it enters the high radio-surface-brightness region of the lobe (Fig.\
\ref{smoothed}), which can be explained in terms of the efficiency of
high-energy particle acceleration in a synchrotron model, but is hard
to explain in terms of the low-energy electrons that would be
responsible for the beamed inverse-Compton model.

A counterjet detection would completely rule out a simple beamed
inverse-Compton model, since the predicted jet-counterjet asymmetry in
such models is very high. The evidence for a counterjet is not strong:
the most convincing region, between about 1 and 2 arcmin from the core
(Fig.\ \ref{smoothed}) contains only about $19 \pm 9$ counts in the
2000 image, or $0.8 \pm 0.4$ nJy. This would imply a jet-counterjet
ratio at this distance from the core around $6_{-2}^{+12}$, using a
matched region to measure the corresponding flux density from the jet. Such a
jet-counterjet ratio is easy to produce with plausible angles to the
line of sight ($\theta < 45$) and moderate speeds ($\beta \sim 0.5$)
in a synchrotron model, but the inverse-Compton models require
$\beta>0.95$ and (for bulk Lorentz factor $\Gamma < 20$) $\theta < 10^\circ$
and for these beaming values the counterjet flux is expected to be
three orders of magnitude below the best estimate of the observed
value. Thus, if the counterjet detection is real, a beamed
inverse-Compton model for the jet is only viable if there is {\it
  also} some X-ray synchrotron radiation from a slow-moving jet region.

\section{Summary and conclusions}

We confirm that the X-ray emission from the lobes of Pic A is most
likely to be due to inverse-Compton scattering of CMB photons, as
found by other workers (Grandi \etal\ 2003, Kataoka \& Stawarz 2005);
the magnetic field strength in the lobes as a whole is inferred to be
a factor 1.6--2 below the equipartition value, as seen in other
sources (Croston \etal\ 2005 and references therein). The inferred
internal pressure implies an external thermal environment no richer
than a moderate group of galaxies, but this is still consistent with
the observations of Faraday depolarization in the lobes (P97).

The large number of counts in the Pic A lobes has allowed us to carry
out a detailed comparison between the radio and X-ray emission. We
find that there are significant differences even at the lowest
frequencies currently accessible; in particular, the inner regions of
the lobes are fainter in the 330-MHz radio while retaining much the
same surface brightness in the X-ray, so that the X-ray/radio ratio is
higher in the inner regions, which also have steep radio spectra.
X-ray spectral fitting allows us to rule out a significant
contribution from extended thermal emission to the X-rays emission in
the inner regions, and nuclear inverse-Compton emission seems unlikely
to be very significant. Simple models in which either the electron
energy spectrum or the magnetic field strength vary as a function of
position in the lobes cannot explain all the observations in the radio
and X-ray, but models in which both vary together as a function of
position can be made consistent with the data. The observations by
Isobe \etal\ (2002) of 3C\,452, which also show an increasing X-ray to
radio ratio in the centre of the source, provide some evidence that
these conclusions can be generalized to the population of FRII sources
as a whole. More deep observations of radio-bright FRII sources are
required to test this.

The region of comparatively bright X-ray emission around the E hotspot
complex (Fig.\ \ref{smoothed}) would require a much lower magnetic
field strength relative to the equipartition value if it were to be
explained by inverse-Compton emission from the hotspot region. Instead
we suggest that this, and similar X-ray emission from regions around
the hotspots of two other low-luminosity FRII radio sources, may be
X-ray synchrotron emission, implying distributed high-energy particle
acceleration in this part of the source.

There is already considerable evidence that the bright jet in the W
lobe is synchrotron in origin: the possible detection of
counterjet-related emission in the E lobe, if real, would rule out a
simple beamed inverse-Compton model for the jet. Unfortunately the
signal to noise in the counterjet region is not good enough for us to
claim a firm detection, but nevertheless our work supports the idea
put forward elsewhere (e.g. Kraft \etal\ 2005)
that X-ray synchrotron is a viable mechanism for emission from FRII
jets in general.

\section*{Acknowledgements}

We are grateful to Andrew Wilson for discussion of Pic A and Rick
Perley for providing the radio maps of the source in electronic form.
We thank an anonymous referee for comments which helped us to improve
the paper. MJH thanks the Royal Society for a research fellowship. The
National Radio Astronomy Observatory is a facility of the National
Science Foundation operated under cooperative agreement by Associated
Universities, Inc. This work is partly based on observations obtained
with {\it XMM-Newton}, an ESA science mission with instruments and
contributions directly funded by ESA Member States and NASA.

\end{document}